\documentclass[a4paper,11pt]{article}
\usepackage[DIV12]{typearea}
\usepackage{longtable}
\usepackage{amsmath}
\usepackage{amssymb}
\usepackage{bbm}
\usepackage[utf8]{inputenc}
\usepackage{pdflscape}
\usepackage{xspace}
\usepackage[caption=false]{subfig}
\usepackage{nicefrac}
\usepackage{graphicx}
\usepackage{cite}

\newcommand{\rep}[1]{\ensuremath\boldsymbol{#1}}

\newcommand{\Z}[1]{\ensuremath{\mathbbm{Z}_{#1}}} 

\newcommand{\U}[1]{\ensuremath{\mathrm{U}(#1)}}

\newcommand{\I}{\mathrm{i}}
\newcommand{\Id}{\mathbbm{1}}

\newcommand{\CP}{\ensuremath{\mathcal{CP}}\xspace}

\usepackage{xcolor}

\definecolor{darkgreen}{HTML}{109930}

\advance \headheight by 3.0truept       
\usepackage[pdftex]{hyperref}
\hypersetup{
    pdftitle = {Unification of Flavor, CP, and Modular Symmetries},
    pdfauthor = {Baur, Nilles, Trautner, Vaudrevange}
}

\begin{document}

\begin{titlepage}

\begin{flushright}
\normalsize{LMU-ASC 02/19}\\
\normalsize{TUM-HEP 1181/19}
\end{flushright}

\vspace*{1.0cm}

\begin{center}
{\Large\textbf{Unification of Flavor, \boldmath $\mathcal{CP}$\unboldmath, and Modular Symmetries}}

\vspace{1cm}

\textbf{Alexander Baur}$^{a}$,
\textbf{Hans~Peter~Nilles}$^{b,c}$, \textbf{Andreas Trautner}$^{d}$, \\ and \textbf{Patrick~K.S.~Vaudrevange}$^{a}$
\\[5mm]
$^a$~Physik Department T75, Technische Universit\"at M\"unchen,\\
James-Franck-Stra\ss e 1, 85748 Garching, Germany\\
\vspace{2mm}
$^b$~Bethe Center for Theoretical Physics and
Physikalisches Institut der Universit\"at Bonn,\\
Nussallee 12, 53115 Bonn, Germany\\
\vspace{2mm}
$^{c}$ Arnold Sommerfeld Center for Theoretical Physics, 
Ludwig-Maximilians-Universit\"at,\\
Theresienstra\ss e 37, 80333 M\"unchen, Germany\\
\vspace{2mm}
$^{d}$ Max-Planck-Institut f\"ur Kernphysik, \\ 
Saupfercheckweg 1, 69117 Heidelberg, Germany
\end{center}

\vspace{1cm}

\vspace*{1.0cm}

\begin{abstract}
Flavor symmetry plays a crucial role in the standard model of particle physics but its origin is 
still unknown. We develop a new method (based on outer automorphisms of the Narain space group) to 
determine flavor symmetries within compactified string theory. A picture emerges where traditional 
(discrete) flavor symmetries, \CP-like symmetries and modular symmetries (like $T$-duality) of 
string theory combine to unified flavor symmetries. The groups depend on the geometry of compact 
space and the geographical location of fields in the extra dimensions. We observe a phenomenon of 
``local flavor groups'' with potentially different flavor symmetries for the various sectors of 
quarks and leptons. This should allow interesting connections to existing bottom-up attempts in 
flavor model building. 
\end{abstract}

\end{titlepage}

\newpage

\section{Introduction}

Traditionally, \CP transformations and flavor symmetries were assumed to be of different origin. 
More recently, however, it was suggested that \CP should be considered as an outer automorphism of 
the flavor group~\cite{Chen:2009gf,Feruglio:2012cw,Holthausen:2012dk,Chen:2014tpa,Trautner:2016ezn}.
It was shown that such a link between flavor and \CP has a natural embedding in a string 
theory framework~\cite{Nilles:2018wex}. In the present paper we show that in string theory, where 
symmetries arise from the geometry of compactified extra dimensions and string selection 
rules~\cite{Kobayashi:2006wq,Nilles:2012cy,Beye:2014nxa}, an even stronger link can be established: 
the \CP and flavor transformations of the low-energy effective theory are unified in a common 
symmetry group. This observation has been made in the generalization of the discussion 
in~\cite{Nilles:2018wex} by including duality symmetries of string theory. In our examples, the 
full unified symmetry (including flavor and \CP) is a combination of the original flavor symmetry 
and the $T$-duality transformations~\cite{Giveon:1988tt,Lauer:1989ax,Lauer:1990tm,Lerche:1989cs,Chun:1989se,Ferrara:1989bc,Ferrara:1989qb}
of the stringy extension. \CP, originally an outer automorphism of the flavor group, now becomes an 
element (inner automorphism) of the unified flavor group. As it contains $T$-duality 
transformations, this unified flavor group depends on the location in moduli space: it is enhanced 
at special points of moduli space (where \CP may be unbroken). At a generic point in moduli space 
only the original flavor symmetry (possibly with \CP as an outer automorphism) is present. This 
allows a connection to the concept of ``local grand unification''~\cite{Forste:2004ie,Buchmuller:2004hv}, 
where the various fields of the standard model of particle physics (quarks, leptons, and Higgs 
bosons) live at different locations in compactified higher dimensions and thus feel different 
subgroups of the unified flavor group. It leads to the flexibility to have different flavor 
symmetries in the various sectors (e.g.\ quark and lepton sectors) of the theory. It also allows a 
connection to model constructions that use $T$-duality transformations, or more general subgroups of 
the modular transformations, as flavor symmetries~\cite{Altarelli:2005yx, deAdelhartToorop:2011re, Feruglio:2017spp,Kobayashi:2018vbk,Kobayashi:2018rad, Penedo:2018nmg, Criado:2018thu,Kobayashi:2018scp,Novichkov:2018ovf,Kobayashi:2018bff,Novichkov:2018nkm,deAnda:2018ecu, Okada:2018yrn, Kobayashi:2018wkl, Novichkov:2018yse}, 
especially for the mixings in the lepton sector.

This unified picture of flavor and \CP is rather common and can be derived through a general 
mechanism that allows a full classification of all flavor symmetries in the given string model. As 
we shall explain in this paper, the mechanism is based on the consideration of outer automorphisms 
of the Narain-lattice construction~\cite{Narain:1985jj,Narain:1986am} of string theory with 
compactified extra dimensions. It is a powerful tool that generalizes previous attempts in the 
search for flavor symmetries. We shall present the mechanism in its general form and explain the 
results in the specific example of the two-dimensional $\Z{3}$ orbifold already discussed in 
ref.~\cite{Nilles:2018wex}. There the flavor group was $\Delta(54)$, and a physical \CP 
transformation had to be a non-trivial outer automorphism of this group. However, due to the 
specific group theoretical structure of $\Delta(54)$ as a ``type I'' group~\cite{Chen:2014tpa}, 
the physical \CP symmetry of the light spectrum was naturally broken by the presence of heavy 
$\Delta(54)$ doublet states. In the generalized picture, $\Delta(54)$ is still the symmetry at a 
generic point in moduli space, but it will be enhanced at specific lines and points of the moduli 
space. Enhancements here include the groups SG(108,17), SG(216,87) and even SG(324,39).\footnote{Here, 
we use the SmallGroup library of \texttt{GAP}~\cite{GAP4} to denote groups.} For these 
enhancements, the \CP transformation of the low-energy spectrum is no longer an outer automorphism, 
but it becomes an element (inner automorphism) of the unified flavor group. Thus, the low-energy \CP 
transformation is conserved at special points, but it will be spontaneously broken at a generic 
point in moduli space.

The main results of our paper originate from the discussion of the outer automorphisms of the 
``Narain space group''~\cite{Narain:1986qm,Narain:1990mw,GrootNibbelink:2017usl}, where we find 
that non-Abelian flavor symmetries, modular symmetries of $T$-duality and \CP have a common origin in 
string theory. Hence, a complete classification of the unified flavor symmetry is possible. First 
we present a warm-up example by the consideration of the outer automorphisms of the geometrical 
space group and then generalize to the discussion of the outer automorphisms of the 
``Narain space group''. The main results of the discussion include:
\begin{itemize}
\item The unification of flavor- and \CP-symmetries,
\item The extension of the traditional flavor group with modular symmetries,
\item A diversification of flavor symmetries to ``local flavor groups'' that depend on the location 
of fields in compactified extra dimensions (and thus could lead e.g.\ to different flavor groups 
for quarks and leptons).
\end{itemize}
We shall apply this to our example of the two-dimensional $\Z{3}$ orbifold and discuss the 
interplay of the original $\Delta(54)$ flavor symmetry and the relevant part of the modular 
transformation of $T$-duality, here given by $\Gamma(3)$ (which is isomorphic to $\mathrm{A}_4$). We explore 
different regions in moduli space and construct the enhanced unified flavor groups.

\section{Outer automorphisms of the space group}
\label{sec:SGAutomorphisms}

\begin{figure*}[t]
\centering{\includegraphics[width=0.4\linewidth]{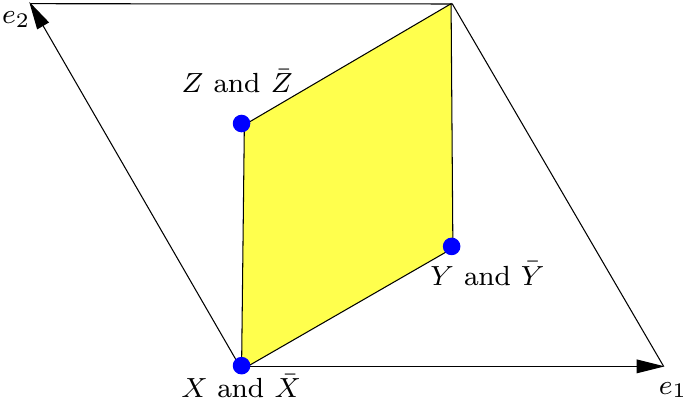}}
\caption{The fundamental domain of the $\mathbbm{T}^2/\Z{3}$ orbifold is depicted in yellow and the 
three inequivalent fixed-points are blue. $(X,Y,Z)$ denote three (left-chiral) twisted strings from the 
first twisted sector, while $(\bar{X},\bar{Y},\bar{Z})$ are three (right-chiral) twisted strings from 
the second twisted sector.}
\label{fig:Z3Orbifold}
\end{figure*}

As a warm-up example we consider the two-dimensional $\Z{3}$ orbifold $\mathbbm{R}^2/S$. For the 
generators of the space group $S$ we can choose $(\theta, 0)$, $(\Id, e_1)$, and $(\Id, e_2)$. The 
vectors $e_1$ and $e_2$ enclose an angle of 120$^\circ$ and have the same length, $|e_1|=|e_2|$. 
They span a two-dimensional lattice which defines the two-torus $\mathbbm{T}^2$. Furthermore, the 
so-called twist $\theta$ is a counter-clockwise 120$^\circ$ rotation matrix with $\theta^3 = \Id$ 
that maps the torus lattice to itself, i.e.\ $\theta\,e_1 = e_2$. In this case, a general space 
group element $g \in S$ can be expressed as 
\begin{equation} \label{eqn:SpaceGroupElement}
g~=~(\theta^k, e\, n) \qquad\text{with}\quad k ~\in~ \{0,1,2\} ~\text{ and }~ n ~\in~\Z{}^2\;.
\end{equation}
Here, the vielbein $e$ contains the two basis vectors $e_1$ and $e_2$ as columns. The element $g$ 
acts on a coordinate $y\in\mathbbm{R}^2$ of the two spatial extra dimensions as 
$y ~\stackrel{g}{\mapsto}~ g\, y ~=~ \theta^k\, y + e\,n$. Consequently, two space group elements 
$(\theta^k, e\, n)$ and $(\theta^\ell, e\, m)$ multiply as 
$(\theta^k, e\, n)\, (\theta^\ell, e\, m) ~=~ (\theta^{k+\ell}, \theta^k e\, m + e\, n)$. Finally, 
the $\mathbbm{T}^2/\Z{3}$ orbifold is defined as a quotient space, i.e.\ $y_1 \sim y_2$ if there 
is a $g \in S$ such that $y_1 = g\, y_2$. This orbifold has three fixed points, i.e.\ points that 
are invariant (up to lattice translations) under the 120$^\circ$ rotation, see 
figure~\ref{fig:Z3Orbifold}. 

Each element $g=(\theta^k, e\,n) \in S$ of the space group describes a boundary condition for a 
closed string on the orbifold~\cite{Dixon:1985jw,Dixon:1986jc}. For example, for a worldsheet boson 
$y(\tau, \sigma)$ of a closed string we impose the boundary condition 
\begin{equation} \label{eqn:BoundaryCondition_g}
y(\tau, \sigma+1) ~=~ g\, y(\tau, \sigma) \quad\Leftrightarrow\quad y(\tau, \sigma+1) ~=~ \theta^k\, y(\tau, \sigma) + e\,n\;.
\end{equation}
$g$ is called the constructing element of the closed string. If $g$ has a fixed point 
$y_\text{f}\in\mathbbm{R}^2$, i.e.\ if $g\,y_\text{f} = y_\text{f}$, the corresponding string 
eq.~\eqref{eqn:BoundaryCondition_g} is localized at $y_\text{f}$ in compactified higher dimensions. 
However, since $h\,y \sim y$ on the orbifold for any $h \in S$, the boundary condition 
eq.~\eqref{eqn:BoundaryCondition_g} and
\begin{equation}
y(\tau, \sigma+1) ~=~ h\,g\,h^{-1} y(\tau, \sigma)
\end{equation}
describe the same closed string on the orbifold. Hence, closed strings on orbifolds are associated 
to conjugacy classes $[g]= \{h\,g\,h^{-1} ~|~ h \in S\}$ of the space group. Then, each conjugacy 
class $[g]$ of the space group $S$ corresponds to a class of boundary conditions and, thereby, to a 
distinct string of the theory. The $\mathbbm{T}^2/\Z{3}$ orbifold has seven conjugacy classes that 
yield massless strings at a generic point in moduli space, i.e.\ for a generic size of the orbifold 
and for generic value of the background $B$-field: The conjugacy class $[(\Id,0)]$ gives the 
trivial boundary condition of massless untwisted strings and twisted strings correspond to the 
conjugacy classes
\begin{subequations}
\begin{align}
X       & ~:~ [(\theta, 0)]\;,  & Y       & ~:~ [(\theta, e_1)]\;,       & Z       & ~:~ [(\theta, e_1+e_2)]\;, \\
\bar{X} & ~:~ [(\theta^2, 0)]\;,& \bar{Y} & ~:~ [(\theta^2, e_1+e_2)]\;, & \bar{Z} & ~:~ [(\theta^2, e_2)]\;,
\end{align}
\end{subequations}
from the first ($\theta$) and second ($\theta^2$) twisted sector, respectively. We choose the 
convention that the string states $(X, Y, Z)$ from the first twisted sector give rise to 
left-chiral degrees of freedom, while $(\bar{X}, \bar{Y}, \bar{Z})$ from the second twisted sector 
yield their right-chiral CPT-conjugates needed to form complete left-chiral superfields. Twisted 
strings are localized at the respective fixed points of their constructing elements. For example, 
the string $X$ with constructing element from the conjugacy class $[(\theta, 0)]$ is localized at 
$y_\text{f} = 0$, see figure~\ref{fig:Z3Orbifold}.

\begin{figure*}[t]
\begin{center}
\subfloat[]{\label{fig:Autoh1}
\includegraphics[width=0.4\linewidth]{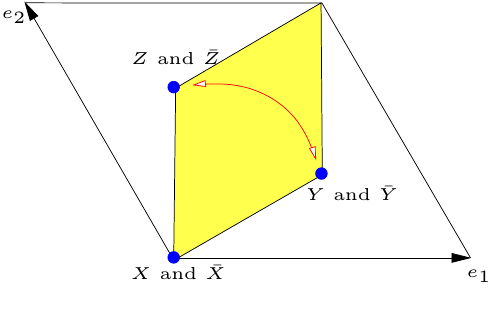}}
\subfloat[]{\label{fig:Autoh2}
\includegraphics[width=0.4\linewidth]{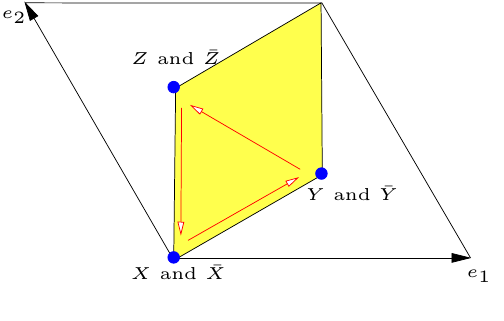}}
\end{center}
\vspace{-0.6cm}
\caption{The $\mathbbm{T}^2/\Z{3}$ orbifold and the actions of the outer automorphisms: (a) the 
180$^\circ$ rotation $\hat{h}_1$ maps $X$ to itself and interchanges $Y$ and $Z$ (and analogously for 
$(\bar{X},\bar{Y},\bar{Z})$) and (b) the $\Z{3}$ translation $\hat{h}_2$ rotates the three twisted 
states. }
\end{figure*}

It is advantageous to change the basis from the so-called coordinate basis to the so-called lattice 
basis. This can be performed for a general space group element, eq.~\eqref{eqn:SpaceGroupElement}, 
as
\begin{equation} \label{eqn:SpaceGroupInLatticeBasis}
\hat{g} ~=~ (e^{-1}, 0)\, (\theta^k, e\,n)\, (e,0) ~=~ (\hat\theta^k, n) ~\in~ \hat{S}\;, \quad\text{where}\quad \hat\theta ~=~ \left(\begin{array}{cc}0 &-1 \\ 1 &-1\end{array}\right)\in\text{GL}(2,\Z{})\;
\end{equation}
is defined via $\hat\theta := e^{-1} \theta\, e$, and $\hat{S}$ denotes the space group in the 
lattice basis. We can now look at the automorphisms of the space group. An inner automorphism,
\begin{equation} 
\hat{g} ~\stackrel{\hat{h}}{\mapsto}~ \hat{h}\,\hat{g}\,\hat{h}^{-1} ~\in~ \hat{S} \quad\text{with}\quad \hat{h}~\in~\hat{S}\;,
\end{equation}
maps each conjugacy class $[\hat{g}]$ to itself and, consequently, each string to itself. 
Furthermore, for a (bosonic) string to be well-defined on an orbifold, it has to be invariant 
under the action of the space group that defines the orbifolded string theory. Hence, inner 
automorphisms act trivially on orbifold-invariant strings. In contrast, an outer automorphism of 
the space group $\hat{S}$ can be described by conjugation of the constructing element with an 
element that is not in $\hat{S}$, i.e.\ there is a transformation 
$\hat{h} := (\hat\sigma, t) \not\in \hat{S}$ such that for all constructing elements 
$\hat{g} \in\hat{S}$ we have
\begin{equation}
\hat{g} ~\stackrel{\hat{h}}{\mapsto}~ \hat{h}\,\hat{g}\,\hat{h}^{-1} ~\stackrel{!}{\in}~ \hat{S}\;.
\end{equation}
Spelled out explicitly, this is equivalent to the following conditions on 
$\hat\sigma\in\text{GL}(2,\Z{})$ and $t$: For each $k \in \{0,1,2\}$ there must be a 
$k' \in \{0,1,2\}$ and $n' \in \Z{}^2$ such that\footnote{%
Here, we have used $\hat{g} = (\hat\theta^k, n) \in\hat{S}$, $\hat{h} = (\hat\sigma, t) \not\in \hat{S}$ 
and we have absorbed $\hat\sigma\, n \in \Z{}^2$ in the definition of $n'$.}
\begin{subequations}\label{eqn:AutoOfSpaceGroup}
\begin{eqnarray}
\hat\sigma\, \hat\theta^k\, \hat\sigma^{-1}                       & \stackrel{!}{=} & \hat\theta^{k'}\;,\\
\left(\Id - \hat\sigma\, \hat\theta^k\, \hat\sigma^{-1}\right)\,t & \stackrel{!}{=} & n'\;.
\end{eqnarray}
\end{subequations}
This is a special case of the general consistency conditions for outer automorphisms~\cite{Trautner:2016ezn}. 
Solutions to these conditions can be written as shifts or rotations, the ones relevant for our 
illustration are generated by the two elements $\hat{h}_i$ given by
\begin{equation}\label{eqn:GeomAutomorphisms}
\hat{h}_1:=(-\Id, 0)\;,\;\; \hat{h}_2:=(\Id, t) \qquad\text{with}\qquad t:=\left(\begin{array}{c} \frac{2}{3} \\ \frac{1}{3}\end{array}\right). 
\end{equation}
These elements of the outer automorphism group have the following geometrical interpretation: 
$\hat{h}_1$ yields a 180$^\circ$ rotation and $\hat{h}_2$ gives a $\Z{3}$ translation, since 
$(\hat{h}_2)^3=(\Id, 3t)$ is an inner automorphism of $\hat{S}$, see figures~\ref{fig:Autoh1} 
and~\ref{fig:Autoh2}. 

Together, the transformations $\hat{h}_1$ and $\hat{h}_2$ generate the permutation group
\begin{equation}
S_3 ~=~ \left\langle \hat{h}_1\,,\hat{h}_2 ~\right|\left. \left(\hat{h}_1\right)^2 = \left(\hat{h}_2\right)^3 = \left(\hat{h}_2\,\hat{h}_1\right)^2 = \Id\right\rangle\;,
\end{equation}
that acts geometrically as all permutations of the three fixed points of the $\Z{3}$ orbifold. If 
one takes into account the $\Z{3}\times\Z{3}$ space group selection rule of the two-dimensional 
$\Z{3}$ orbifold~\cite{Hamidi:1986vh,Ramos-Sanchez:2018edc}, the combined flavor symmetry results 
as~\cite{Kobayashi:2006wq}
\begin{equation}
\Delta(54) ~=~ S_3 \ltimes \left(\Z{3}\times\Z{3}\right)\;.
\end{equation}
Note that in ref.~\cite{Kobayashi:2006wq} the $S_3$-symmetry had been postulated from geometrical 
considerations while here we have deduced it from the outer automorphisms of the space group. 
As a remark, our approach eq.~\eqref{eqn:AutoOfSpaceGroup} is similar to the identification of 
flavor symmetries in complete intersection Calabi-Yau manifolds~\cite{Lukas:2017vqp,Candelas:2017ive}.
Still, this is not the full picture. To obtain the complete flavor group of string theory we have 
to analyze the outer automorphisms of the Narain space group. This Narain approach will also reveal 
``non-geometric'' symmetries that are not accessible in the geometrical approach: For example, also 
the $\Z{3}\times\Z{3}$ space group selection rules (as discussed in our warm-up example here) will 
be part of the outer automorphisms of the Narain space group. In the more general picture there 
will be enhanced flavor symmetries, some of which originate from the modular symmetries of string 
theory and appear to be non-universal in moduli space.

\section{Outer automorphisms of the Narain space group}
\label{sec:NarainAutomorphisms}

In this section we extend the discussion of outer automorphisms of the geometrical space group to 
the Narain space group. It turns out that the outer automorphisms of the Narain space group give 
rise to the full (non-Abelian) unified flavor symmetry of the theory: it includes (i) possible 
permutation symmetries of the various fixed points and sectors of the orbifold, (ii) the space 
group selection rule of strings splitting and joining while moving on the surface of the orbifold, 
(iii) the target-space modular symmetries of $T$-duality and (iv) \CP-like transformations.\footnote{%
A \CP-like transformation is a transformation that acts like a physical \CP transformation on some 
but not all states of a theory~\cite{Chen:2014tpa}.} In fact, the total resulting flavor symmetry 
depends on the precise value of the K\"ahler and complex structure moduli, i.e.\ colloquially 
speaking, on the region in moduli space. In more detail, for certain shapes/sizes of the orbifold 
and for certain values of the background $B$-field the (non-Abelian) flavor symmetry gets enhanced.

We focus mainly on the bosonic string coordinates and consider a symmetric $\mathbbm{T}^D/\Z{K}$ 
orbifold with $\Z{K}$ twist $\Theta$ in the Narain formulation, see appendix~\ref{app:NarainLattice}. 
In this formulation the $D$-dimensional compactified string coordinates $y$ are separated into $D$ 
right- and $D$ left-moving degrees of freedom, $y_\text{R}$ and $y_\text{L}$, collectively denoted 
by $Y = (y_\text{R}, y_\text{L})^\text{T}$. Now, in analogy to the geometrical construction 
discussed in section~\ref{sec:SGAutomorphisms}, $Y$ is compactified on a $2D$-dimensional torus 
defined by a $2D$-dimensional Narain lattice with vielbein $E$, composed out of basis vectors $E_i$ 
for $i=1,\dots,2D$, and a metric with $(D,D)$ signature $\eta=\text{diag}(-\Id,\Id)$. The $\Z{K}$ 
Narain space group $S_\text{Narain}$ is defined by its generators $(\Theta, 0)$, $(\Id, E_i)$ for 
$i=1,\ldots,2D$. A general element $g\in S_\text{Narain}$ reads
\begin{equation}
g ~=~ (\Theta^k, E\,\hat{N})\;, \quad\text{with}\quad \Theta = \left(\begin{array}{cc}\theta_\text{R} & 0 \\0 & \theta_\text{L}\end{array}\right), \quad k\in\{0,\ldots,K-1\}\; \quad\text{and}\quad \hat{N}\in\Z{}^{2D}\;,
\end{equation}
where $\hat{N}$ contains the $D$ winding and $D$ Kaluza-Klein (KK) quantum numbers of the string. 
The $\Z{K}$ Narain orbifold can be described as an extension of the geometrical construction 
discussed in section~\ref{sec:SGAutomorphisms} from coordinates $y$ to right- and left-movers $Y$. 
Therefore, one identifies
\begin{equation}
Y ~\sim~ g\,Y ~=~ \Theta^k\,Y + E\,\hat{N}\;.
\end{equation}
A symmetric $\Z{K}$ orbifold is obtained under the assumption that the $2D$-dimensional rotation 
matrix $\Theta$ acts left-right-symmetrically on the $2D$-dimensional Narain lattice, i.e.\ 
$\theta_\text{R} = \theta_\text{L}=\theta$ for symmetric $\Z{K}$ orbifolds. Finally, we change the 
basis to the lattice basis (denoted by hatted quantities),
\begin{equation}\label{eq:HatTheta}
\hat{g} ~=~ (E^{-1}, 0)\, (\Theta^k, E\,\hat{N})\, (E,0) ~=~ (\hat\Theta^k, \hat{N}) ~\in~ \hat{S}_\text{Narain}\;, \quad\text{where}\quad \hat\Theta ~:=~ E^{-1} \Theta\, E\;.
\end{equation}

Similar to the geometric space group, a conjugacy class $[\hat{g}]$ of the Narain space group 
defines a class of boundary conditions and, thereby, gives rise to a distinct string. Strings on 
orbifolds have to be invariant under inner automorphisms of the Narain space group 
$\hat{S}_\text{Narain}$. Hence, inner automorphisms of $\hat{S}_\text{Narain}$ act trivially on 
orbifold-invariant strings. In contrast, outer automorphisms of $\hat{S}_\text{Narain}$ correspond 
to the symmetries of the full (bosonic) string theory on orbifolds. Outer automorphisms of the 
Narain space group $\hat{S}_\text{Narain}$ are given by transformations 
$\hat{h}:=(\hat\Sigma, \hat{T}) ~\not\in~ \hat{S}_\text{Narain}$ which act on each element 
$\hat{g}\in\hat{S}_\text{Narain}$ such that 
\begin{equation}
\hat{g} ~\stackrel{\hat{h}}{\mapsto}~ \hat{h}\,\hat{g}\,\hat{h}^{-1} ~\stackrel{!}{\in}~ \hat{S}_\text{Narain}\;.
\end{equation}
Spelled out explicitly, this is equivalent to a set of consistency conditions requiring that for 
each $k$ there must be an $k' \in \{0,\ldots,K-1\}$ and $\hat{N}' \in \Z{}^{2D}$ such that 
$\hat\Sigma\in\text{GL}(2D,\Z{})$ and
\begin{subequations}\label{eqn:AutoOfNarainSpaceGroup}
\begin{eqnarray}
\hat\Sigma\, \hat\Theta^k\, \hat\Sigma^{-1}                             & \stackrel{!}{=} & \hat\Theta^{k'}\;,\label{eqn:AutoOfNarainSpaceGroup1}\\
\left(\Id - \hat\Sigma\, \hat\Theta^k\, \hat\Sigma^{-1}\right)\,\hat{T} & \stackrel{!}{=} & \hat{N}'\;,      \label{eqn:AutoOfNarainSpaceGroup2}
\end{eqnarray}
\end{subequations}
in analogy to eq.~\eqref{eqn:AutoOfSpaceGroup}. The translational part can be fractional, 
i.e.\ $\hat{T}\not\in\Z{}^{2D}$, and $\hat\Sigma$ may not be a $\Z{K}$ rotation, i.e.\ 
$\hat\Sigma\neq \hat\Theta^\ell$ for $\ell=1,\ldots,K-1$. In addition to the consistency 
conditions, $\hat\Sigma$ must be a modular transformation (i.e.\ it must satisfy 
$\hat{\Sigma}^\text{T} \hat\eta\, \hat{\Sigma} = \hat\eta$) in order to be a symmetry of the Narain 
lattice.

As a solution to these conditions one can find a set of generators of the outer automorphism group 
of the form
\begin{equation}
\left\{ (\hat\Sigma_1, 0),\; (\hat\Sigma_2, 0),\;\ldots,\; (\Id, \hat{T}_1),\; (\Id, \hat{T}_2),\;\ldots \right\}\;,
\end{equation}
i.e.\ the outer automorphism group can be generated by pure rotations 
$(\hat\Sigma_i, 0)\not\in \hat{S}_\text{Narain}$ and pure translations 
$(\Id, \hat{T}_j)\not\in \hat{S}_\text{Narain}$ -- roto-translations $(\hat\Sigma, \hat{T})$ are 
not needed as generators\footnote{Assuming we had a non-trivial roto-translation 
$(\hat\Sigma, \hat{T})$ as a solution to eq.~\eqref{eqn:AutoOfNarainSpaceGroup}. Then also 
$(\hat\Sigma, 0)$ and $(\Id, \hat{T})$ are solutions.}. 

Finally, it is instructive to transform the matrix $\hat\Sigma$ back to the coordinate basis such 
that
\begin{equation}
\hat\Sigma ~=~ E^{-1} \Sigma\, E \;, \quad\text{subject to the condition}\quad \Sigma^\text{T} \eta\,\Sigma ~\stackrel{!}{=}~ \eta\;.
\end{equation}
A flavor symmetry transformation $\Sigma$ should leave 
$(p_\text{R})^2$ and $(p_\text{L})^2$ invariant (and consequently the right- and left-moving string 
masses). Thus, we have to demand
\begin{equation}
\Sigma ~\stackrel{!}{=}~ \begin{pmatrix} \sigma_\text{R}&0\\0&\sigma_\text{L}\end{pmatrix} \quad\text{where}\quad \sigma_\text{R}, \sigma_\text{L} ~\in~ \text{O}(D) \quad\Leftrightarrow\quad \Sigma^\text{T} \Sigma ~=~ \Id\;.
\end{equation}
Demanding this condition on the outer automorphism $(\Sigma,0)$ leaves the compactification moduli 
invariant. Such a transformation belongs to the traditional flavor symmetry.

Let us now specialize to traditional flavor symmetries in two compactified dimensions $D=2$. 
Since in this case $\sigma_\text{R}$ and $\sigma_\text{L}$ are two-dimensional orthogonal matrices, $\sigma_\text{R}, \sigma_\text{L} \in \text{O}(2)$, 
each of them can be uniquely parametrized by one angle, $\alpha_\text{R}$ and $\alpha_\text{L}$, respectively. For example
\begin{equation}
\sigma_\text{R}(\alpha_\text{R}) ~=~ \begin{pmatrix} \cos(\alpha_\text{R}) & \mp \sin(\alpha_\text{R})\\ \sin(\alpha_\text{R}) & \pm \cos(\alpha_\text{R})  \end{pmatrix}\;,
\end{equation} 
where the different signs correspond to a rotation or a reflection, respectively. As a consequence 
of~\eqref{eqn:AutoOfNarainSpaceGroup1}, for $\Z{K}$ orbifolds with $K\neq 2$ in $D=2$ the matrices 
$\sigma_\text{R}$ and $\sigma_\text{L}$ have to have the same determinant,
\begin{equation}
\text{det}(\sigma_\text{R}) ~=~ \text{det}(\sigma_\text{L})\;.
\end{equation}
Hence, the matrices $\sigma_\text{R}$ and $\sigma_\text{L}$ describe either both rotations or both 
reflections, labeled by subscripts ``rot.'' or ``refl.'', respectively. In both cases, $\Sigma$ can 
be symmetric (i.e.\ $\sigma_\text{R} = \sigma_\text{L}$ thus $\alpha = \alpha_\text{R} = \alpha_\text{L}$) or 
asymmetric (i.e.\ $\sigma_\text{R} \neq \sigma_\text{L}$ thus $\alpha_\text{R} \neq \alpha_\text{L}$) 
and we denote the corresponding $\hat\Sigma$-matrix in the lattice basis by $\hat{S}(\alpha)$ or 
$\hat{A}(\alpha_\text{R}, \alpha_\text{L})$, respectively. Consequently, going back to the lattice 
basis, the outer automorphisms $(\hat\Sigma,0)$ fall into four categories, where $\hat\Sigma$ can be 
either
\begin{equation}\label{eq:fourCases}
\hat{S}_\text{rot.}(\alpha)\;,\;\hat{S}_\text{refl.}(\alpha)\;,\;\hat{A}_\text{rot.}(\alpha_\text{R}, \alpha_\text{L})\;,\quad\text{or}\quad \hat{A}_\text{refl.}(\alpha_\text{R}, \alpha_\text{L})\;.
\end{equation}
Reflections are obviously of order 2, i.e.\ 
$\hat{S}_\text{refl.}(\alpha)^2 = \hat{A}_\text{refl.}(\alpha_\text{R}, \alpha_\text{L})^2 = \Id$. 
On the other hand, rotations must map the four-dimensional Narain lattice to itself. Thus, the 
order of four-dimensional rotations is restricted to the values
\begin{equation}
\{1,2,3,4,5,6,8,10,12\}\;,
\end{equation}  
using the Euler-$\phi$ function~\cite{schwarzenberger1980n}.

In general, the $4 \times 4$ matrices $\hat\Sigma=E^{-1} \Sigma\, E$ obtained from 
eq.~\eqref{eq:fourCases} are integer matrices only for special values of the moduli that 
parametrize the vielbein $E$ of the Narain lattice. Thus, the traditional flavor symmetry obtained 
from the outer automorphism group of the Narain space group depends on the value of the moduli.

\section[Flavor symmetries of the Z3 orbifold]{\boldmath Flavor symmetries of the $\Z{3}$ orbifold\unboldmath}
\label{sec:FlavorFromNarainAutomorphisms}

In this section we shall use the method based on the automorphisms of the Narain space group to 
determine the traditional flavor symmetry of the $\mathbbm{T}^2/\Z{3}$ orbifold. We will see that the 
non-Abelian flavor symmetry $\Delta(54)$ is a subgroup of the unified flavor group: $\Delta(54)$
will be enhanced in certain regions of moduli space by the modular symmetries of the underlying 
string theory. 

To obtain the symmetric $\Z{3}$ orbifold of the $(2,2)$-dimensional Narain lattice $\Gamma$ we 
choose a special Narain vielbein $E$ (see appendix~\ref{app:NarainLattice}) by setting
\begin{equation}\label{eqn:NarainLatticeChoice}
e ~=~ R\begin{pmatrix}1&-\frac{1}{2} \\[2pt] 0 & \phantom{0} \frac{\sqrt{3}}{2}\end{pmatrix}\;
\quad \text{and}\quad B ~=~ b\,\alpha'\begin{pmatrix}0&1\\[2pt]-1&0\end{pmatrix}\;.
\end{equation}
The real parameters $b$ and $r:=\nicefrac{R^2}{\alpha'}$ are combined into the K{\"a}hler modulus 
$T$ and the complex structure modulus $U$ of the two-torus, i.e.\
\begin{equation}
T ~=~ b + \I\, \frac{\sqrt{3}}{2}\,r \quad\text{and}\quad U ~=~ \exp\left(\frac{2\pi\I}{3}\right)\;.
\end{equation}
We see that the complex structure modulus $U$ is frozen and the K{\"a}hler modulus $T$ parametrizes 
the $(b,r)$-moduli space of the symmetric $\Z{3}$ Narain orbifold in $D=2$. This means that the 
Narain lattice $\Gamma$ can be deformed by freely choosing the K{\"a}hler modulus $T$ while the 
$\Z{3}$ symmetry of $\Gamma$ is kept intact. Now, the (left-right-symmetric) $\Z{3}$ Narain 
orbifold can be specified by the Narain space group defined by the elements
\begin{equation}\label{eq:Z3HatTheta}
\hat{g} ~=~ (\hat\Theta^k, \hat{N}) ~\in~ \hat{S}_\text{Narain}\;, \quad\text{with}\quad \hat\Theta ~=~ E^{-1} \Theta\, E ~=~ \begin{pmatrix}\hat\theta & 0 \\ 0 & \hat\theta^{-T}\end{pmatrix}\;,
\end{equation}
where the two-dimensional twist matrix $\hat\theta = e^{-1} \theta\, e$ is given in 
eq.~\eqref{eqn:SpaceGroupInLatticeBasis}. Due to the choice of the Narain vielbein $E$ specified in 
eq.~\eqref{eqn:NarainLatticeChoice} and appendix~\ref{app:NarainLattice}, the Narain twist 
$\hat\Theta$ in the lattice basis is a symmetry of the Narain lattice 
$\hat\Theta^\text{T} \hat\eta\, \hat\Theta = \hat\eta$, as necessary. The symmetries of the 
$(2,2)$-dimensional Narain lattice are isomorphic to $\text{SL}(2,\Z{})_T \times \text{SL}(2,\Z{})_U$ 
and $B \mapsto -B$ (together with $G_{12} \mapsto -G_{12}$~\cite{Lerche:1989cs}). After the 
orbifolding the complex structure modulus $U$ is fixed, thus $\text{SL}(2,\Z{})_U$ is broken. 
Therefore, we can focus on the remaining $\text{SL}(2,\Z{})_T$ modular symmetry and on the 
$\CP$-like transformation $B \mapsto -B$.

Under $\text{SL}(2,\Z{})_T$ modular transformations untwisted and twisted strings transform 
non-trivially~\cite{Lauer:1989ax}. Acting on massless strings only, the modular group turns out to 
be $\mathrm{T}'$, i.e.\ the double covering group of $A_4\simeq \Gamma(3)$~\cite{Lerche:1989cs}. 
Combined with the transformation $B \mapsto -B$ the group gets further enhanced to 
$\text{GL}(2, 3)$ (i.e.\ $\text{SG}(48, 29)$).
 
As described in section~\ref{sec:NarainAutomorphisms}, the relevant outer automorphisms of the 
Narain lattice are described by pure rotations $(\hat\Sigma_i,0)$ from the modular symmetries 
$\hat\Sigma_i \in \text{SL}(2,\Z{})_T$ and pure translations $(\Id,\hat{T}_j)$ of the four-dimensional 
Narain lattice that fulfill the consistency condition~\eqref{eqn:AutoOfNarainSpaceGroup}. 
Generally, the transformations $\hat\Sigma_i$ act non-trivially on the $T$-modulus. However, at 
some specific points in moduli space some transformations $\hat\Sigma_i$ might leave the vacuum 
expectation value (VEV) $\langle T\rangle$ of the K{\"a}hler modulus invariant. Then, 
$\hat\Sigma_i$ is an element of the traditional flavor symmetry at the point $\langle T\rangle$ in 
$T$-moduli space. If $\hat\Sigma_i$ leaves $\langle T\rangle$ invariant we get 
$\Sigma_i \in \text{O}(2)\times\text{O}(2)$ and these rotations fall into the four categories 
specified in eq.~\eqref{eq:fourCases}: symmetric and asymmetric rotations and symmetric and 
asymmetric reflections. In the following we shall discuss these specific cases. We shall only 
present the results here and provide the more technical derivation in a future publication, which 
will include a derivation of the transformation behavior of twisted and untwisted strings under the action of the 
outer automorphisms of the Narain space group, and a full discussion of the origin of the modular 
symmetry $\text{GL}(2, 3)$.

\subsection[Generic Point in <T>]{\boldmath Generic Point in $\langle T\rangle$\unboldmath} 
\label{sec:GenericPoint}
 
At a generic point $\langle T\rangle$ in the $T$-moduli space, the outer automorphisms that leave 
the K\"ahler modulus $T$ invariant can be generated by two translations $\mathrm{A}$ and 
$\mathrm{B}$ and a symmetric rotation $\mathrm{C}$ given by $\hat{S}_\text{rot.}(\pi) = -\Id$ in 
eq.~\eqref{eq:fourCases}, i.e.
\begin{equation}\label{eqn:NarainAutomorphisms}
\mathrm{A} ~:=~ (\Id, \hat{T}_1)\;,\;  \mathrm{B} ~:=~ (\Id, \hat{T}_2)  \quad\text{with}\quad \hat{T}_1 ~:=~ \begin{pmatrix}\frac{1}{3}\\\frac{2}{3}\\0\\0\end{pmatrix} \;,\; \hat{T}_2 ~:=~ \begin{pmatrix}0\\0\\\frac{1}{3}\\\frac{1}{3}\\\end{pmatrix}\;,\quad\text{and}\quad \mathrm{C} ~:=~ (-\Id, 0)\;.
\end{equation}
The automorphism $\mathrm{A}$ shifts the winding number and $\mathrm{B}$ shifts the KK number. 
Comparing eq.~\eqref{eqn:NarainAutomorphisms} to the generators of the outer automorphisms of the 
geometrical space group $\hat{h}_1 = (-\Id, 0)$ and $\hat{h}_2 = (\Id, t)$, listed in 
eq.~\eqref{eqn:GeomAutomorphisms}, we can identify the correspondences
\begin{equation}
\mathrm{A}^2 ~\leftrightarrow~ \hat{h}_2 \quad\text{and}\quad \mathrm{C} ~\leftrightarrow~ \hat{h}_1\;.
\end{equation}
The $\Z{3}$ Narain outer automorphism $\mathrm{B}$ is not accessible in the geometrical case. 
Acting with $\mathrm{B}$ on untwisted and twisted strings, we observe that $\mathrm{A^2B^2AB}$ and 
$\mathrm{B}$ give rise to the $\Z{3}\times\Z{3}$ point and space group selection 
rules~\cite{Hamidi:1986vh,Ramos-Sanchez:2018edc}. Altogether, $\mathrm{A}$, $\mathrm{B}$, and 
$\mathrm{C}$ generate the flavor symmetry $\Delta(54)$ at a generic point in moduli space from the 
outer automorphisms of the Narain space group. This should be compared to the result in our warm-up 
example discussed earlier, where only the $S_3$ subgroup could be obtained from the outer automorphisms of the 
geometrical space group.

\begin{figure}[t]
\centering{\includegraphics[width=0.8\linewidth]{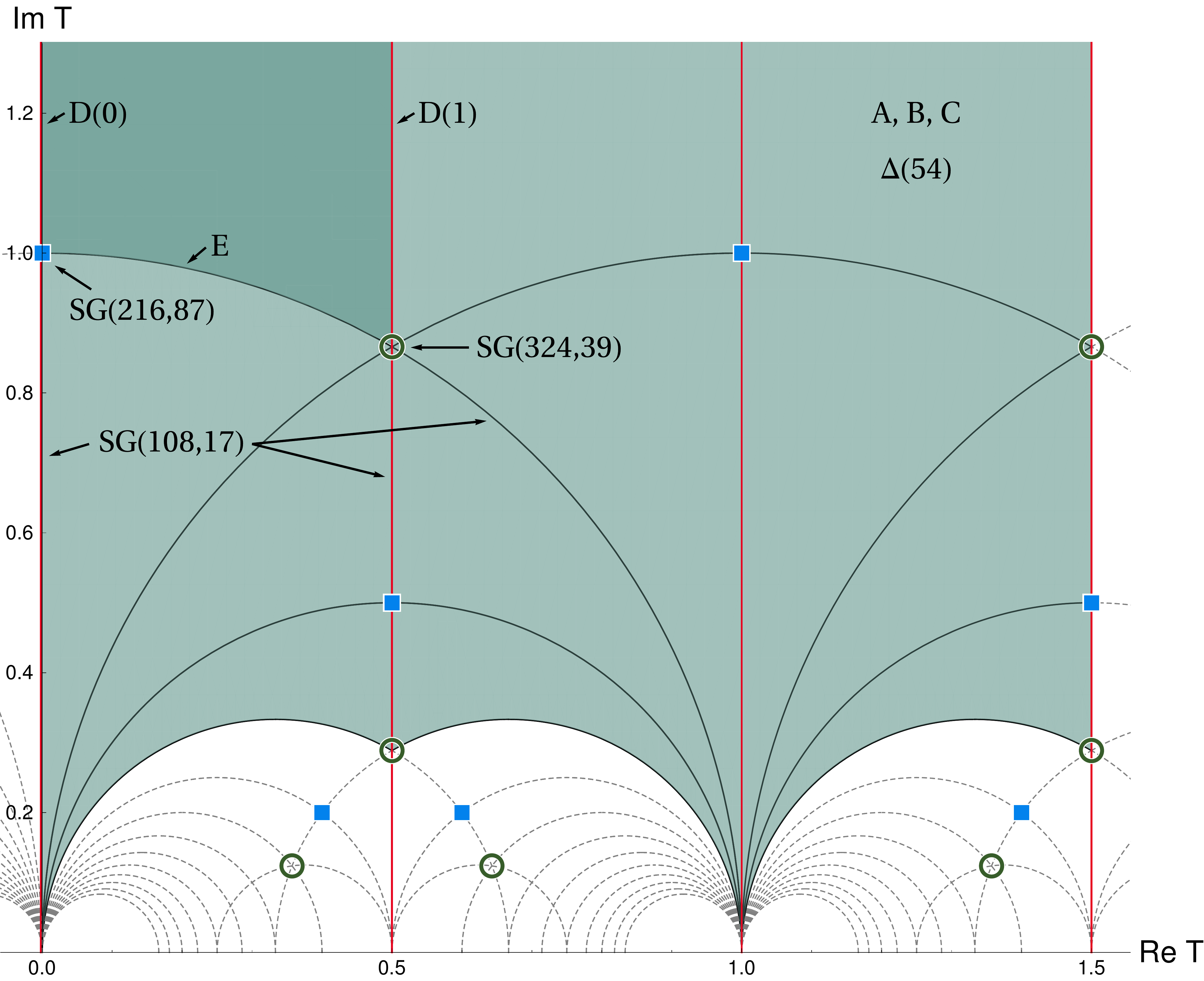}}
\caption{Points and curves of flavor symmetry enhancement in the moduli space 
\mbox{$T=b + \I\, \nicefrac{\sqrt{3}}{2}\,r$}. The dark teal area is the fundamental domain of 
$\text{SL}(2,\Z{})_T$, while the light teal area is the one of $\mathrm{T}'$ (for both, modulo $B \mapsto -B$). On 
the vertical red lines and on the black semicircles the flavor group $\Delta(54)$ gets enhanced 
to $\text{SG}(108,17)$, see sections~\ref{sec:HalfIntB} and~\ref{sec:semicircle}. When two lines 
intersect, i.e.\ points marked by blue squares, the symmetry is further enhanced to 
$\text{SG}(216, 87)$, as discussed for $(b,r)=(0,\nicefrac{2}{\sqrt{3}})$ in section~\ref{sec:SG216}. 
When three lines intersect, i.e.\ points marked by small green circles, the flavor group is even 
further enhanced to $\text{SG}(324, 39)$, see section~\ref{sec:SG324} for the point $(b,r)=(\nicefrac{1}{2},1)$.}
 \label{Fig:modplane}
\label{modplot}
\end{figure}

\subsection[Special B-field with b=integer/2]{\boldmath Special $B$-field with $b=\frac{1}{2}\times\text{integer}$ \unboldmath}
\label{sec:HalfIntB}

Let us now show that for generic radii $r$ but quantized values of the $B$-field, $b=\nicefrac{n_B}{2}$ with $n_B\in\Z{}$,
the flavor symmetry gets enhanced. Consider the left-right-symmetric reflective outer automorphism transformation 
\begin{equation}
\mathrm{D}(n_B)~:=~(\hat{S}_\text{refl.}(\nicefrac{2\pi}{6}),0)\;, \quad \text{with} \quad
\hat{S}_\text{refl.}(\alpha=\nicefrac{2\pi}{6})~=~ \left(\begin{array}{cccc}
 1   &  0   & 0 &  0 \\
 1   & -1   & 0 &  0 \\
-n_B &  n_B & 1 &  1 \\
n_B &  0   & 0 & -1 
\end{array}\right)\;,
\end{equation}
where $\hat{S}_\text{refl.}$ has been introduced in equation~\eqref{eq:fourCases}. 
This transformation has a residual moduli dependence (here in the form of $n_B$), which is a 
possibility already noted at the end of section~\ref{sec:NarainAutomorphisms}. Nonetheless, 
$\mathrm{D}(n_B)^2=(\Id,0)$ as expected for a reflection. However, $\mathrm{D}(n_B)$ is a symmetry 
of the Narain lattice only if $n_B$ takes integer values (cf.\ appendix~\ref{app:NarainLattice}).
The $T$-modulus transforms under $\mathrm{D}(n_B)$ as
\begin{equation}
T ~\mapsto~ n_B - \overline{T}\;.
\end{equation}
Hence, the VEV $\langle T\rangle$ is invariant
\begin{equation}
\langle T\rangle ~\mapsto~ n_B - \overline{\langle T\rangle} ~=~ \langle T\rangle \quad\text{for}\quad \langle T\rangle ~=~ \frac{n_B}{2} + \I\, \frac{\sqrt{3}}{2}\,\langle r\rangle\;.
\end{equation}
Therefore, at regions in moduli space where $b=\nicefrac{n_B}{2}$ with $n_B\in\Z{}$ there appears 
an unbroken $\Z{2}$ transformation generated by $\mathrm{D}(n_B)$. This enhances the $\Delta(54)$ 
flavor symmetry to SG(108,17), see figure~\ref{Fig:modplane}.

The six-dimensional representation $\rep{6}$ of SG(108,17) acts faithfully on the six twisted 
strings $(X, Y, Z, \bar{X}, \bar{Y}, \bar{Z})$. The $\rep{6}$ branches into 
$\rep{3}\oplus\rep{\bar3}$ of the $\Delta(54)$ subgroup, implying that $\mathrm{A}$, $\mathrm{B}$, 
and $\mathrm{C}$ only act separately in the barred and unbarred subspaces. On the contrary, 
$\mathrm{D}(n_B)$ acts as an interchange of $(X,Y,Z)$ and their \CP-partners 
$(\bar{X},\bar{Y},\bar{Z})$ (possibly with $n_B$-dependent phases). This is backed-up by that fact 
that, geometrically, $\mathrm{D}(n_B)$ acts as a reflection on the axis perpendicular to $e_2$ and, 
consequently, it corresponds to complex conjugation in the extra dimensions~\cite{Strominger:1985it}. 
Thus, from a $\Delta(54)$ point of view, the $\Z{2}$ transformation $\mathrm{D}(n_B)$ acts as a 
\CP-like transformation. In the previous work~\cite{Nilles:2018wex} outer automorphisms of the 
flavor group $\Delta(54)$ were considered as candidates for \CP-like symmetries. And indeed, the 
$\Z{2}$ transformation $\mathrm{D}(n_B)$ is contained in the outer automorphism group $S_4$ of 
$\Delta(54)$. Altogether, we see that at the specific lines $b=\nicefrac{n_B}{2}$ in moduli space 
the flavor- and \CP-symmetries are unified into a single symmetry group.  

Moreover, deflecting the VEV of the $T$-modulus away from the symmetry-enhanced point
\begin{equation}
\langle T\rangle ~=~ \frac{n_B}{2} + \I\, \frac{\sqrt{3}}{2}\,\langle r\rangle \quad \text{to} \quad \langle T'\rangle ~=~ \langle T\rangle + \delta T \quad\text{with}\quad \text{Re}(\delta T) \neq 0\;,
\end{equation}
induces a spontaneous symmetry breaking of SG(108,17) to $\Delta(54)$. This shows that the unified 
flavor symmetry, and more specifically the \CP-like transformation $\mathrm{D}(n_B)$, can be broken 
spontaneously.

\subsection{Black Circles}
\label{sec:semicircle}

There are more regions in the $T$-moduli space with an enhanced flavor symmetry. For example, on 
the semicircle 
\begin{equation}\label{eq:semicircle}
|T|^2 ~=~ 1 \quad\text{with}\quad \text{Im}(T) ~>~ 0\;,
\end{equation}
a specific left-right-asymmetric reflection outer automorphism 
$\hat{A}_\text{refl.}\left(\alpha_\text{R},\alpha_\text{L}\right)$ becomes an element of the 
symmetry group. Here, $\alpha_\text{R}$ and $\alpha_\text{L}$ depend on $T$ precisely in such a way 
as to ensure that the additional symmetry generated by 
\begin{equation}\label{eq:E}
\mathrm{E} ~:=~ (\hat{A}_\text{refl.1},0)\;, \quad\text{with}\quad \hat{A}_\text{refl.1} ~:=~ \begin{pmatrix}0&0&0&1\\0&0&1&1\\-1&1&0&0\\1&0&0&0\end{pmatrix}\;,
\end{equation}
holds everywhere on the semi-circle. 

Similar to the case in section~\ref{sec:HalfIntB}, amending the generators $\mathrm{A}$, 
$\mathrm{B}$, $\mathrm{C}$ by the $\Z{2}$ transformation $\mathrm{E}$ enhances the $\Delta(54)$ 
flavor symmetry to SG(108,17) everywhere on the semicircle. Despite being isomorphic, the two 
SG(108,17) groups here and in section~\ref{sec:HalfIntB} are not identical, i.e.\ they are 
different extensions of $\Delta(54)$. Analogous enhancements happen on the other black semicircles 
depicted in figure~\ref{Fig:modplane}.

\subsection[Special point at b=0 and r=2/sqrt(3)]{\boldmath Special point at $b=0$ and $r=\nicefrac{2}{\sqrt{3}}$\unboldmath}
\label{sec:SG216}

Let us now consider a case where two lines meet, for example the point 
$(b,r)=(0,\nicefrac{2}{\sqrt{3}})$ in the $T$-moduli space, marked by a blue square in 
figure~\ref{Fig:modplane}. At this point, the unbroken generators, in addition to the usual 
ones of $\Delta(54)$, are $\mathrm{D}(0)$ and $\mathrm{E}$. The total symmetry group at this point 
then can simply be computed as the closure of all generators, and the result is $\text{SG}(216, 87)$. 

We remark that other, apparently independent, transformations might be conserved at this point as 
well. For example, the left-right-asymmetric 4-fold rotation 
$\hat{A}_\text{rot.}(\nicefrac{2\pi}{4},\nicefrac{-2\pi}{4})$. However, none of these additional 
transformations is independent of the transformations above as all of them are already contained in 
$\text{SG}(216, 87)$.

\subsection[Special point at b=1/2 and r=1]{\boldmath Special point at $b=\nicefrac{1}{2}$ and $r=1$\unboldmath}
\label{sec:SG324}

Finally, we consider a point in the $T$-moduli space where three lines meet, for example 
$(b,r)=(\nicefrac{1}{2},1)$. There, we identify the following flavor symmetries:\footnote{At the 
point $(b,r)=(\nicefrac{1}{2},1)$ there is an additional $\U{1}^2$ gauge symmetry enhancement.} 
$\mathrm{A}$, $\mathrm{B}$, and $\mathrm{C}$ originate from the generic case, $\mathrm{D}(1)$ 
appears on the vertical line at $b=\nicefrac{1}{2}$, while $\mathrm{E}$ is the additional flavor 
symmetry on the semicircle $|T|^2 = 1$. In addition, for the semicircle with center $(b,r)=(1,0)$ 
we identify the unbroken reflection
\begin{equation}
\mathrm{F}~:=~(\hat{A}_\text{refl.2},0)\;, \quad\text{with}\quad 
\hat{A}_{\text{refl.2}} ~:=~ \left(\begin{array}{cccc}
 1 &  0 & 0 & -1 \\
 1 & -1 &-1 & -1 \\
 0 &  0 & 1 &  1 \\
 0 &  0 & 0 & -1 
\end{array}\right)\;.
\end{equation}
This transformation is not independent of the others, as $\mathrm{F}=\mathrm{C\,E\,D(1)\,E}$. Also 
all other possible additional symmetries out of the set~\eqref{eq:fourCases}, which may be 
envisaged at this specific point, turn out to be dependent. Thus, the total flavor symmetry at 
$(b,r)=(\nicefrac{1}{2},1)$ can be computed as the closure of 
$\{\mathrm{A}, \mathrm{B}, \mathrm{C}, \mathrm{D}(1), \mathrm{E}\}$ and the result is 
$\text{SG}(324, 39)$, see figure~\ref{Fig:modplane}.

\section{Conclusions and Outlook}
\label{sec:conclusions}

In the present paper we have given a unified description of \CP- and flavor-symmetries in string 
theory. This was possible through the development of a new tool to obtain the full classification 
of flavor symmetries. It is based on the investigation of outer automorphisms of the Narain space 
group of compactified string theory. Apart from the traditional flavor symmetries (as discussed 
in~\cite{Kobayashi:2006wq,Nilles:2012cy,Beye:2014nxa}), this approach includes string dualities as 
well. The unified flavor group has the peculiar property that it is non-universal in the moduli 
space of compactified extra dimensions. Different regions (points or lines) in moduli space might 
enjoy enhanced flavor symmetries. This allows the unification of \CP-transformations within the 
unified flavor symmetries. The spontaneous breakdown of \CP is then controlled by the vacuum 
expectation values of the moduli fields. We have illustrated this in a specific example based on 
the $\Z{3}$ orbifold. There we identify the traditional universal flavor group $\Delta(54)$ at 
generic points in moduli space, with enhancements to unified flavor symmetries SG(108,17), 
SG(216,87) up to SG(324,39). The enhanced groups are pretty large although our analysis only 
considered a two-dimensional compactified space, whereas in string theory we have altogether six 
additional compact space dimensions. 

The picture discussed here makes contact to the previous work on \CP-violation described 
in ref.\cite{Nilles:2018wex}. The phenomenological implications are still valid here, but we 
have gained a new perspective in the sense that the explicit breakdown of \CP in~\cite{Nilles:2018wex} 
can now be understood as a spontaneous breakdown of \CP within the unified picture. In addition, 
the new perspective presented here offers novel directions for flavor model building. As we find 
different flavor symmetries at different points in moduli space (in particular in six compact 
dimensions), fields that live at different locations in moduli space feel a different amount of 
flavor symmetry. Applied to the standard model of particle physics, this could explain, for 
example, why the observed flavor structure of quarks and leptons is so different. This is 
reminiscent of the concept of ``local grand unification''~\cite{Forste:2004ie,Buchmuller:2004hv} 
where we can identify different enhanced gauge groups at different ``geographical'' 
locations~\cite{Nilles:2014owa} in compact extra dimensions.

The enhanced unified flavor groups are pretty large (especially in the realistic case of six
compact dimensions) and allow flexibility for a step-wise breakdown through Wilson 
lines~\cite{Ibanez:1986tp,Ibanez:1987xa,Ibanez:1987pj} and the vacuum expectation values of the 
moduli of compact space. This could lead to a different flavor- and \CP-structure for the various 
sectors of the standard model like up- or down-quarks, charged leptons or neutrinos. Such a scheme 
would share similarities with flavor constructions discussed recently~\cite{Hagedorn:2018gpw,Hagedorn:2018bzo}. 
It would also connect to bottom-up constructions that use duality transformations for models of 
mixing in the quark and especially the lepton sector~\cite{Feruglio:2017spp,Kobayashi:2018vbk,Kobayashi:2018rad, Penedo:2018nmg, Criado:2018thu,Kobayashi:2018scp,Novichkov:2018ovf,Kobayashi:2018bff,Novichkov:2018nkm,deAnda:2018ecu, Okada:2018yrn, Kobayashi:2018wkl, Novichkov:2018yse}, 
although there are two major differences between these studies and our present point of view. The 
first one comes from the fact that in our picture the flavor symmetries are a hybrid combination of 
traditional and modular discrete symmetries, whereas in~\cite{Feruglio:2017spp,Kobayashi:2018vbk,Kobayashi:2018rad, Penedo:2018nmg, Criado:2018thu,Kobayashi:2018scp,Novichkov:2018ovf,Kobayashi:2018bff,Novichkov:2018nkm,deAnda:2018ecu, Okada:2018yrn, Kobayashi:2018wkl, Novichkov:2018yse} 
the flavor symmetries are assumed to be completely contained within the modular group. The second 
difference is a consequence of the fact that in the string theory picture the K\"ahler potential 
(and thus the superpotential as well) transforms non-trivially under modular 
transformations~\cite{Lauer:1989ax,Lauer:1990tm,Lerche:1989cs,Chun:1989se,Ferrara:1989bc,Ferrara:1989qb} 
in contrast to the assumption of the papers mentioned above. We shall elaborate on the details of 
these differences in a future publication. 

The present discussion shows that string theory naturally leads to a rich and flexible flavor 
structure that could explain many different aspects of flavor- and \CP-symmetry in the standard 
model. It is worthwhile to go ahead with future research in that direction, both from the bottom-up 
and top-down perspective.

\section*{Acknowledgments}
H.P.N.\ thanks Dieter L\"ust and the Arnold Sommerfeld Center at LMU Munich for hospitality and support. 
P.V.\ is supported by the Deutsche Forschungsgemeinschaft (SFB1258). 

\appendix

\section{Narain lattice}
\label{app:NarainLattice}
 
We consider string compactifications on a $(D,D)$-dimensional Narain lattice $\Gamma$ and its 
(symmetric) $\mathbbm{T}^D/\Z{K}$ orbifold in $D$ extra dimensions, following the conventions of 
ref.~\cite{GrootNibbelink:2017usl}. In this case the Narain lattice is a $2D$-dimensional lattice. 
It is defined by $2D$ basis vectors $E_{i=1,\dots,2D}$, which we combine into a $(2D \times 2D)$-dimensional vielbein 
matrix $E$. In more detail, the Narain lattice $\Gamma$ can be defined by a torus compactification 
of right- and left-moving bosonic string coordinates $y_\text{R}$ and $y_\text{L}$, respectively, 
i.e.
\begin{equation}\label{eqn:NarainLattice}
Y ~\sim~ Y + E\,\hat{N} \quad\text{with}\quad Y ~=~ \left(\begin{array}{c}y_\text{R}\\y_\text{L}\end{array}\right) ~\in~\mathbbm{R}^{2D} \quad\text{and}\quad \hat{N} ~=~ \left(\begin{array}{c}n\\m\end{array}\right) ~\in~\Z{}^{2D}\;,
\end{equation}
where $n\in\Z{}^D$ and $m\in\Z{}^D$ are the winding and Kaluza-Klein quantum numbers, respectively. 
The string coordinates $y$ and their $T$-duals $\tilde{y}$ are related by
\begin{equation}\label{eqn:CoordinatesYAndTildeY}
\begin{pmatrix}y\\ \tilde{y}\end{pmatrix} ~=~ \frac{1}{\sqrt{2}} \begin{pmatrix}\Id & \Id\\-\Id & \Id\end{pmatrix}\, \begin{pmatrix}y_\text{R}\\ y_\text{L}\end{pmatrix}\;.
\end{equation}
For the string theory to have a modular invariant partition function, the vielbein $E$ has to span 
an even self-dual lattice with signature $(D,D)$ -- called the Narain lattice $\Gamma$. Hence, in 
the absence of Wilson lines the vielbein $E$ can be parametrized as
\begin{equation}\label{eqn:vielbeinE}
E ~=~ \dfrac{1}{\sqrt{2}}\,
\begin{pmatrix}
\dfrac{e^{-\text{T}}}{\sqrt{\alpha'}}\,(G-B) &           -\sqrt{\alpha'}\,e^{-\text{T}} \\[0.4cm]
\dfrac{e^{-\text{T}}}{\sqrt{\alpha'}}\,(G+B) & \phantom{-}\sqrt{\alpha'}\,e^{-\text{T}}
\end{pmatrix}\;,
\end{equation}
where $e$ is the geometrical vielbein of the $D$-dimensional torus $\mathbbm{T}^D$ with metric $G:=e^\text{T}e$, $e^{-\text{T}}$ denotes the transposed inverse of $e$, 
$\alpha'$ denotes the Regge slope, and $B=-B^\mathrm{T}$ is the anti-symmetric $B$-field. From eq.~\eqref{eqn:vielbeinE} 
it follows that 
\begin{equation}\label{eqn:NarainCondition}
E^\text{T} \eta\, E ~=~ \hat\eta\;, \quad\text{where}\quad \eta ~:=~ \begin{pmatrix} -\Id & 0\\ 0 & \Id\end{pmatrix} \quad\text{and}\quad \hat\eta ~:=~ \begin{pmatrix} 0 & \Id\\ \Id & 0\end{pmatrix}\;.
\end{equation}
A transformation of the vielbein $E \mapsto E'~= E\, \hat{M}^{-1}$ is a symmetry of the Narain lattice $\Gamma$ iff
\begin{equation}
\hat{M} ~\in~ \text{GL}(2D,\Z{}) \qquad\text{and}\qquad \hat{M}^\text{T} \hat\eta\, \hat{M} ~=~ \hat\eta\;.
\end{equation}

\bibliography{Orbifold}
\bibliographystyle{OurBibTeX}
\end{document}